\documentclass[aps,prb,reprint]{revtex4-2}

\usepackage[dvipdfmx]{graphicx}
\usepackage{amsmath}
\usepackage{amssymb}
\usepackage[dvipdfmx]{color}
\usepackage{siunitx}
\usepackage{multirow}

\usepackage{comment}

\renewcommand{\i}{\mathrm{i}}
\newcommand{\e}{\mathrm{e}}
\renewcommand{\d}{\mathrm{d}}

\newcommand{\kk}{\mathbf{k}}
\newcommand{\pp}{\mathbf{p}}

\newcommand{\Aa}{\mathbf{A}}
\newcommand{\Ee}{\mathbf{E}}
\newcommand{\Dd}{\mathbf{D}}
\newcommand{\Jj}{\mathbf{J}}
\newcommand{\Pp}{\mathbf{P}}
\newcommand{\rr}{\mathbf{r}}
\newcommand{\vv}{\mathbf{v}}
\newcommand{\ee}{\mathbf{e}}
\newcommand{\Rr}{\mathbf{R}}

\newcommand{\Eg}{E_\mathrm{g}}

\newcommand{\xxi}{\boldsymbol{\xi}}
\newcommand{\Ww}{\boldsymbol{\Omega}}

\newcommand{\Tr}{\operatorname{Tr}}

\begin{document}

\title{
Unified theory of the photovoltaic Hall effect by field- and light-induced Berry curvatures
}

\date{\today}
\author{Yuta Murotani}
\affiliation{The Institute for Solid State Physics, The University of Tokyo, Kashiwa, Chiba 277-8581, Japan}

\author{Tomohiro Fujimoto}
\affiliation{The Institute for Solid State Physics, The University of Tokyo, Kashiwa, Chiba 277-8581, Japan}

\author{Ryusuke Matsunaga}
\affiliation{The Institute for Solid State Physics, The University of Tokyo, Kashiwa, Chiba 277-8581, Japan}

\begin{abstract}
Photovoltaic Hall effect, i.e., generation of a photocurrent perpendicular to the bias electric field, is an interesting platform of Berry curvature engineering by external fields. 
Floquet engineering aims at generation of light-induced Berry curvature associated with topological phase transition in solids, which may manifest itself as a light-induced anomalous Hall effect. 
However, recent studies have pointed out a larger contribution by momentum asymmetry of photocarriers, termed a field-induced circular photogalvanic effect. 
Except for numerical studies, the two mechanisms have been described by different theoretical frameworks, hindering a coherent understanding. 
Here, we develop a unified theory of the photovoltaic Hall effect capable of describing both mechanisms on an equal footing. 
We reveal that the bias electric field alters the interband transition dipole moment and transition energy, both contributing to the field-induced circular photogalvanic effect in nonmagnetic materials. 
These effects are governed by an electric field-induced Berry curvature and the shift vector coupled to bias field, respectively. 
A resonant enhancement of the transverse photocurrent is found in GaAs owing to the topological character of the valence band. 
We also clearly distinguish the anomalous Hall effect by light-dressed states within the density matrix calculation using the length gauge. 
Our theory unifies a number of nonlinear optical processes in a physically transparent way and presents a geometric picture of the third-order nonlinear response under light and bias fields, shedding new light on Berry curvature engineering.
\end{abstract}

\maketitle

\section{Introduction}

Berry curvature associated with a geometric phase is essential in determining the properties of materials with broken time-reversal ($\mathcal{T}$) or inversion ($\mathcal{I}$) symmetry \cite{Resta2000,Xiao2010,Vanderbilt2018}.
In the $\mathcal{T}$-breaking case, {Berry curvature leads to} the anomalous Hall effect (AHE), in which an electric current generates a transverse voltage without an external magnetic field \cite{Karplus1954,Chang2008,Nagaosa2010}.
A close relationship between geometry and topology has guided the discovery of a quantized version of the AHE \cite{Haldane1988,Yu2010,Chang2013}.
By contrast, in the $\mathcal{I}$-breaking case, Berry curvature leads to the circular photogalvanic effect (CPGE), where circularly polarized light generates a photocurrent known as an injection current \cite{Sipe2000,Hosur2011,Orenstein2021,Morimoto2023}. 
The possibility of a quantized CPGE has been discussed for topological Weyl semimetals \cite{Juan2017,Avdoshkin2020,Jankowski2024}.
Even in nonmagnetic and centrosymmetric materials, Berry curvature engineering by external fields can be devised to induce these effects, because circularly polarized light and an electric field can be used to break $\mathcal{T}$ and $\mathcal{I}$, respectively.

Existing studies on the photovoltaic Hall effect, {i.e.}, generation of a photocurrent perpendicular to a bias electric field, have focused on the function of circularly polarized light that breaks $\mathcal{T}$ to induce the AHE.
The mechanisms include Floquet engineering, which modifies the band structure to generate a transient Berry curvature \cite{Oka2009,Chan2016,McIver2020,Chen2021,Yoshikawa2022,Hirai2023,Hirai2024,Day2024,Cao2024}, and {the} optical selection rule, which allows interband excitation of electrons carrying a nonvanishing Berry curvature \cite{Bakun1984,Miah2007,Yin2011,Virk2011,Yu2012,Xiao2012,Mak2014,Okamoto2014,Priyadarshi2015,Fujimoto2024,Murotani2024,Xiao2012,Mak2014}.
However, recent studies have shown that momentum asymmetry of photoexcited carriers contributes more significantly to the Hall current, when the light and bias fields are applied simultaneously \cite{Sato2019a,Sato2019b,Nguyen2021,Murotani2023,Fujimoto2025}.
Although this mechanism has been conceived to be a field-induced CPGE caused by $\mathcal{I}$ breaking under a bias electric field \cite{Murotani2023,Fujimoto2025}, its relation with the field-induced Berry curvature has not been addresed yet.
Deeper insight into the field-induced CPGE is required to clarify the geometric aspects of light--matter interaction.
Furthermore, it is strongly demanded to synthesize the contributions by light-dressed Floquet states and excited carriers into a single theoretical framework, since the exising theories have described them seperately using different techniques.

\begin{figure}[t]
\centering
\includegraphics[width=\columnwidth]{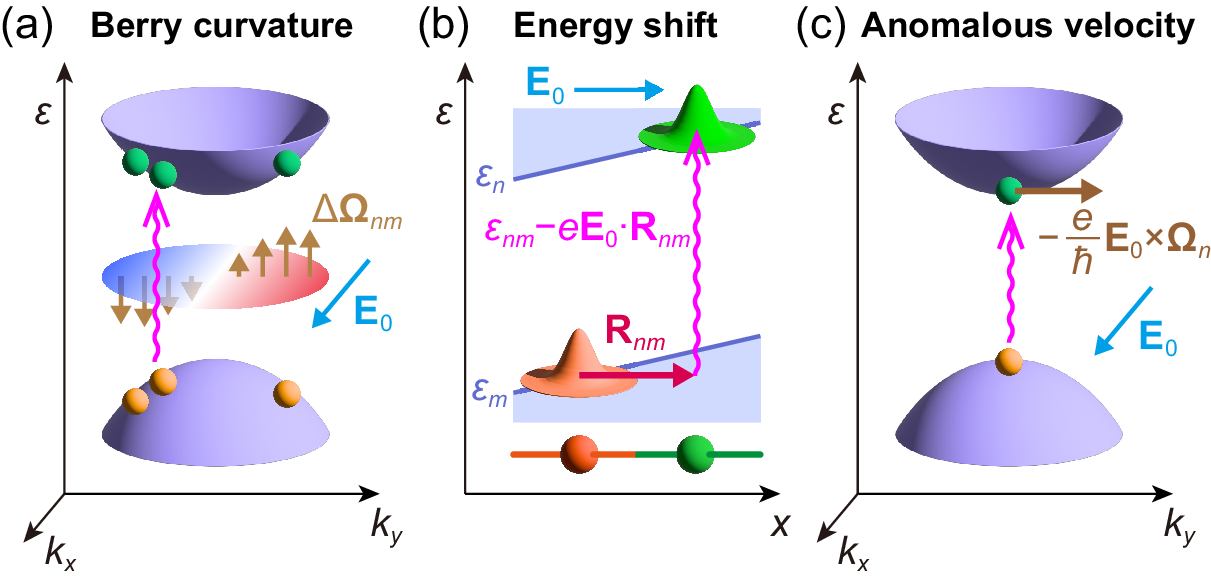}
\caption{(a) Band-resolved Berry curvature $\Delta\Ww_{nm}$ generated by a bias electric field $\Ee_0$, which causes a momentum asymmetry of photoexcited carriers.
(b) Field-induced energy shift of interband transitions, arising from the potential gradient corresponding to the shift vector $\Rr_{nm}(\ee)$.
(c) Anomalous velocity of photoexcited carriers generated by the Berry curvature $\Ww_n$.
}\label{fig:schematic}
\end{figure}

In this study, we develop a general theory of the photovoltaic Hall effect in the third-order nonlinear regime, treating the field-induced CPGE and the light-induced AHE on an equal footing.
First, we elucidate the physical processes underlying the field-induced CPGE, which are summarized in Fig. \ref{fig:schematic}.
We establish the relationship between the field-induced CPGE and field-induced Berry curvature for the first time, as illustrated in Fig. \ref{fig:schematic}(a).
We also find an additional contribution to the Hall current by a change in the transition energy, which is illustrated in Fig. \ref{fig:schematic}(b).
This effect involves the shift vector, another geometric quantity that measures the spatial shift of electron clouds accompanying interband transitions \cite{Sipe2000}. 
The field-induced Berry curvature and energy shift, combined with the anomalous velocity of photoexcited carriers [Fig. \ref{fig:schematic}(c)], describe the photovoltaic Hall effect in nonmagnetic insulators.
We demonstrate our results for a massive Dirac electron system and a typical semiconductor, GaAs.
The topological character of the valence band in the latter leads to divergence in all the mechanisms, explaining resonance peaks in the photon energy dependence of the photovoltaic Hall effect observed recently \cite{Fujimoto2025}.
Next, we calculate the contribution by light-dressed states in metallic systems using the same density matrix method. 
Our result not only agrees with the prediction by Floquet theory but also reveals a close link between light-induced AHE and optical rectification, both related to a light-induced Berry connection.
These findings provide a unified picture of the photovoltaic Hall effect and underline its geometric aspect.

This paper is organized as follows.
In Sec. II, we derive the field-induced Berry curvature and prove its role in the field-induced CPGE.
Section III reveals another contribution arising from a coupling between the shift vector and electric field.
Anomalous velocity of photocarriers is briefly outlined in Sec. IV to obtain a comprehensive picture of the photovoltaic Hall effect in insulators.
Photon energy dependence of each mechanism is analyzed in Sec. V, revealing an enhancement of the transverse current by a degeneracy point in band structure.
Section VI completes the theory by presenting the contribution from light-dressed states and shows its relationship with optical rectification. 
Section VII compares our theory with previous studies.
Finally, Sec. VIII summarizes our findings.

\section{Field-induced Berry curvature}

We begin with a brief review of the CPGE.
In a nonmagnetic insulator, it is described by \cite{Sipe2000,Juan2017,Watanabe2021}
\begin{align}
\dot{\Jj}_{\mathrm{inj}}& = \frac{e^3}{2\hbar^2}\sum_{\kk}\sum_n\sum_{m>n}f_{nm}(\boldsymbol{\Gamma}_{nm}\cdot\Ww_{nm})(\nabla_{\kk}\omega_{nm}),\label{eq:Jinj_1}
\end{align}
where $e < 0$ denotes {the} electron charge, 
$f_{nm}$ is the difference in occupancy between the initial ($m$) and final ($n$) states, and 
$\omega_{nm}$ is the transition frequency.
{$\kk$ dependence of each quantity is omitted for brevity.}
Equation \eqref{eq:Jinj_1} {consists of} velocity ($\nabla_{\kk}\omega_{nm}$) and {the number} ($\propto f_{nm}\boldsymbol{\Gamma}_{nm}\cdot\Ww_{nm}$) {of photoexcited carriers}. The latter depends on two material characteristics:
transition energy $\hbar\omega_{nm}$, which is included in  $\boldsymbol{\Gamma}_{nm} = 2\pi\i\Ee_1(\omega)\times\Ee_1^*(\omega)\delta(\omega-\omega_{nm})$, 
and band-resolved Berry curvature $\Ww_{nm}=\i\xxi_{nm}\times\xxi_{mn}$, derived from the Berry connection matrix $\xxi_{nm} = \langle u_n|\i\nabla_{\kk}|u_m\rangle$.
The well-known Berry curvature $\Ww_n = \nabla_{\kk}\times\boldsymbol{\xi}_{nn}$ is equal to the sum of band-resolved one, $\sum_{m\neq n}\Ww_{nm}$.

Appearance of $\Ww_{nm}$ in Eq. \eqref{eq:Jinj_1} can be understood as follows.
The transition matrix element for an interband transition $m\to n$ is given by $M_{nm}(\ee)=(\ee\cdot\xxi_{nm})(\ee^*\cdot\xxi_{mn})$,
where $\ee$ denotes the polarization vector of light.
This equation enables us to relate the band-resolved Berry curvature to the circular dichroism,
\begin{align}
\Omega_{nm}^z&=M_{nm}(\ee_{\mathrm{R}})-M_{nm}(\ee_{\mathrm{L}}),\label{eq:CD_2}
\end{align}
where $\ee_{\mathrm{L,R}}=(1/\sqrt{2})(1,\pm\i,0)$ is the polarization vector for left- and right-circularly polarized light propagating in the $z$ direction.
The band-resolved Berry curvature thus quantifies the circular dichroism at each $\kk$ point.
When this circular dichroism is combined with the inversion symmetry breaking, a photocurrent is generated in a direction dependent on the helicity of incident light, which is nothing but the CPGE.
In two-band systems, the band-resolved Berry curvature is equivalent to the total Berry curvature, i.e., $\Ww_1=\Ww_{12}$ and $\Ww_2=\Ww_{21}$.
In this case, CPGE is described purely by the total Berry curvature, which makes geometrical and topological consideration easy \cite{Juan2017,Orenstein2021}.
In systems with more energy bands, the band-resolved Berry curvature deviates from the mathematical definition of curvature unless it is summed over bands, hindering simple link with geometry or topology.
Nevertheless, $\Ww_{nm}$ can be considered to quantify how much the $m$-th band bends the subspace of $n$-th band.

{When a static electric field $\Ee_0$ is applied, the interband Berry connection acquires a correction $\Delta\xxi_{nm} = C_{nm}\Ee_0$, where
\begin{align}
C_{nm}^{ab}& = \frac{\i e}{\hbar}D^a\left(\frac{\xi_{nm}^b}{\omega_{nm}}\right)+\frac{e}{\hbar}\sum_{l\neq n,m}\left(\frac{\xi_{nl}^a\xi_{lm}^b}{\omega_{lm}}-\frac{\xi_{nl}^b\xi_{lm}^a}{\omega_{nl}}\right),\label{eq:Cnm_1}
\end{align}
is the interband Berry connection polarizability, and $D^aO_{nm} = (\partial^a-\i\xi_{nn}+\i\xi_{mm})O_{nm}$ is a generalized \cite{Aversa1995} or covariant \cite{Fregoso2018} derivative.
Among the Berry connection polarizability, intraband components ($n = m$) were first recognized in the theory of nonlinear transport \cite{Gao2014,Liu2021,Liu2022}.
Equation \eqref{eq:Cnm_1} extends this concept to the interband components ($n\neq m$).
The modified Berry connection leads to a correction $\Delta\Ww_{nm} = \Pi_{nm}\Ee_0$ to the band-resolved Berry curvature, governed by another polarizability
\begin{align}
\Pi_{nm}^{ad}& = \i\sum_{bc}\epsilon^{abc}T_{nm}^{bcd}.\label{eq:Pinm_1}
\end{align}
Here, $\epsilon^{abc}$ is the Levi--Civita symbol, and
$T_{nm}^{bcd} = \xi_{nm}^bC_{mn}^{cd}+C_{nm}^{bd}\xi_{mn}^c$.
The electric field-induced Berry curvature adds a correction to Eq. \eqref{eq:Jinj_1},}
\begin{align}
\dot{\Jj}_{\mathrm{inj,dip}}& = \frac{e^3}{2\hbar^2}\sum_{\kk}\sum_n\sum_{m>n}f_{nm}(\boldsymbol{\Gamma}_{nm}\cdot\Delta\Ww_{nm})(\nabla_{\kk}\omega_{nm}),\label{eq:Jinjdip_1}
\end{align}
{which can be nonzero even in centrosymmetric materials.}
Although we presented equations for a non-degenerate case, systems with degenerate bands can be treated similarly by appropriately choosing the bases {(see Appendix \ref{sec:SVD})}.
{A more general expression applicable to any choice of bases is given by Eq. \eqref{eq:Jinjdip_2} in Appendix \ref{sec:J2}.}

\begin{figure}[t]
\centering
\includegraphics[width=\columnwidth]{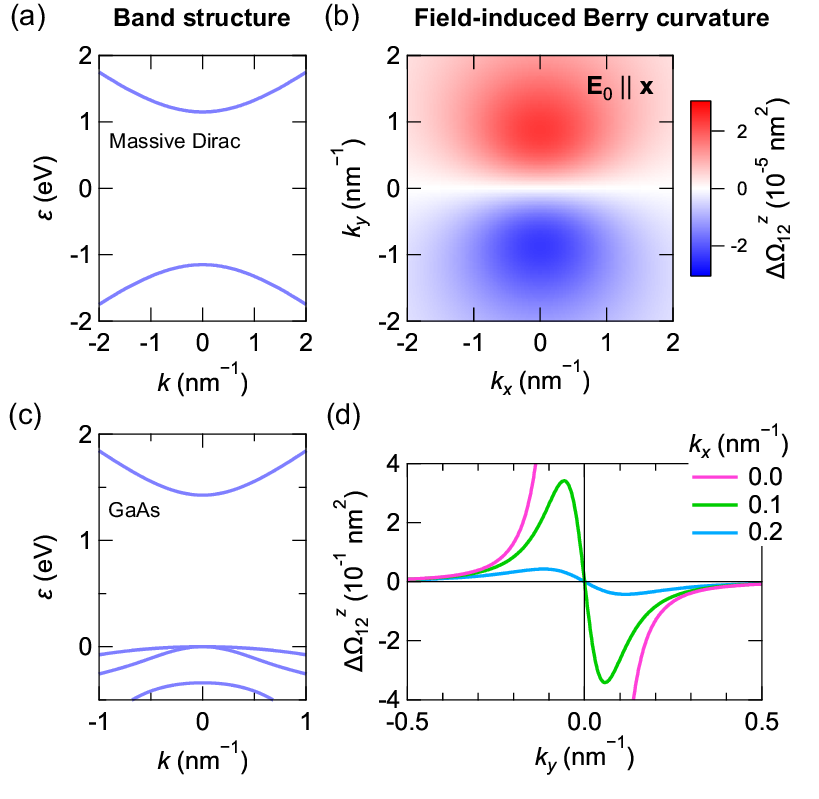}
\caption{(a) Band structure of a massive Dirac electron system. 
(b) Field-induced correction to the $z$ component of the band-resolved Berry curvature in a massive Dirac electron system.
A bias field $E_0 = 10$ kV/cm in the $x$ direction is assumed.
(c) Band structure of GaAs described using an eight-band Kane model.
(d) Field-induced correction to the $z$ component of the band-resolved Berry curvature between CB and HH in GaAs.
A bias field $E_0 = 10$ kV/cm in the $x$ direction is assumed.
$k_z$ is fixed at 0 nm$^{-1}$, and $k_x$ is varied as indicated in the figure.
The curve for $k_x = 0$ nm$^{-1}$ diverges at $k_y = 0$ nm$^{-1}$.
}\label{fig:FI-BC}
\end{figure}

As a simple example, {we consider a} massive Dirac electron system shown in Fig. \ref{fig:FI-BC}(a).
The dispersion relation is given by $\epsilon_{1,2} = \pm\Lambda$, where $\Lambda = \sqrt{\Delta^2+\hbar^2v^2k^2}$.
We set $2\Delta = 2.3$ eV and $v = 1\times10^6$ m/s, considering a lead halide perovskite \cite{Volosniev2023}.
This group of materials is expected to be useful in spintronics \cite{Lu2024}. 
The field-induced Berry curvature takes the form of $\Delta\Ww_{12} = (e\hbar^4v^4/2\Lambda^5)\kk\times\Ee_0$ \cite{Liu2022}.
Figure \ref{fig:FI-BC}(b) illustrates the $z$ component of $\Delta\Ww_{12}$ on the $k_z = 0$ plane, assuming a bias field in the $x$ direction.
A sign change occurs across the horizontal axis, generating a photocurrent in the $y$ direction following Eq. \eqref{eq:Jinjdip_1}. 

A semiconductor GaAs exhibits a unique behavior despite its prototypical appearance.
Figure \ref{fig:FI-BC}(c) shows the band structure approximated using a first-order Kane model. 
The conduction (CB), heavy-hole (HH), light-hole (LH), and split-off bands were labeled 1 to 4, respectively.
We calculated the field-induced Berry curvature near the $\Gamma$ point using the $\kk\cdot\pp$ perturbation method. 
While $\Delta\Ww_{14}\propto e\Ee_0\times\kk$ exhibits a structure similar to that of a massive Dirac system, 
$\Delta\Ww_{12} = -\Delta\Ww_{13} = (3e/2\Eg k^4)\Ee_0\times\kk$ diverges at the zone center, as shown in Fig. \ref{fig:FI-BC}(d).
This divergence originates from a strong mixing of energetically close HH and LH bands under $\Ee_0$, enforced by degeneracy at the $\Gamma$ point. 
Notably, this degeneracy entails monopole charges $Q = \pm3$ in the momentum space, which serves as a source of the Berry curvature through $\nabla_{\kk}\cdot\Ww_n = 2\pi\delta(\kk)Q$ for $n = 2,3$ \cite{Murakami2003}. 
The sign of $Q$ depends on the helicity of electrons associated with the total angular momentum $j = 3/2$.
As such, the valence band in GaAs intrinsically bears a topological character, which is hidden in thermal equilibrium owing to the cancellation of the Berry curvature by equally occupied electronic states.
The divergence of $\Delta\Ww_{12}$ and $\Delta\Ww_{13}$ can be considered as an expression of such a nontrivial topology at the top of the valence band under a bias electric field.
We demonstrate later that this divergence can be observed through the resonant enhancement of the field-induced CPGE.
A previous study on semiconductors proposed a similar mechanism for the photovoltaic Hall effect \cite{Dai2007}.
However, the existence of a Fermi surface was required, whereas our $\Jj_{\mathrm{inj,dip}}$ appears irrespective of carrier doping.

\section{Field-induced energy shift}

The second correction to the CPGE is described by
\begin{align}
\dot{\Jj}_{\mathrm{inj,ene}}& = \frac{e^3}{2\hbar^2}\sum_{\kk}\sum_n\sum_{m>n}f_{nm}(\nabla_{\kk}\omega_{nm})\nonumber\\
&\quad\times\left(\Delta\bar{\omega}_{nm}\frac{\partial\boldsymbol{\Gamma}_{nm}}{\partial\omega_{nm}}\cdot\Ww_{nm}\right),\label{eq:Jinjene_1}
\end{align}
where
\begin{align}
\Delta\bar{\omega}_{nm}& = \frac{|\ee\cdot\xxi_{nm}|^2\Delta\omega_{nm}(\ee)-(\ee\to\ee^*)}{|\ee\cdot\xxi_{nm}|^2-(\ee\to\ee^*)},
\end{align}
is the weighted average of $\Delta\omega_{nm}(\ee) = -e\Ee_0\cdot\Rr_{nm}(\ee)/\hbar$ and $\Delta\omega_{nm}(\ee^*)$.
{Here,} $\Rr_{nm}(\ee) = \xxi_{nn}-\xxi_{mm}-\nabla_{\kk}\operatorname{arg}(\ee\cdot\xxi_{nm})$ is the well-known shift vector for interband transition $m\to n$ \cite{Sipe2000}.
$\hbar\Delta\omega_{nm}(\ee)$ is equal to the difference in the electrostatic potential between the initial and final states at $\kk$, and $\hbar\Delta\omega_{nm}(\ee^*)$ is that at $-\kk$.
A previous study proposed that Rashba-type spin-orbit coupling should modify the band structure under a bias field to generate a Hall current \cite{Yu2013}, similar to the picture suggested by Eq. \eqref{eq:Jinjene_1}.
However, $\Rr_{nm}(\ee)$ and $\hbar\Delta\omega_{nm}(\ee)$ depend on the polarization vector $\ee$ of light, indicating that the band structure under a bias field should appear different when observed using light with different polarization{s}.
A description based on Rashba-type spin-orbit coupling is valid only when the shift vector is independent of light polarization.
{Materials with degenerate bands can be dealt with by Eq. \eqref{eq:Jinjene_3} in Appendix \ref{sec:J2} or by single value decomposition in Appendix \ref{sec:SVD}.}

\begin{figure}[t]
\centering
\includegraphics[width=\columnwidth]{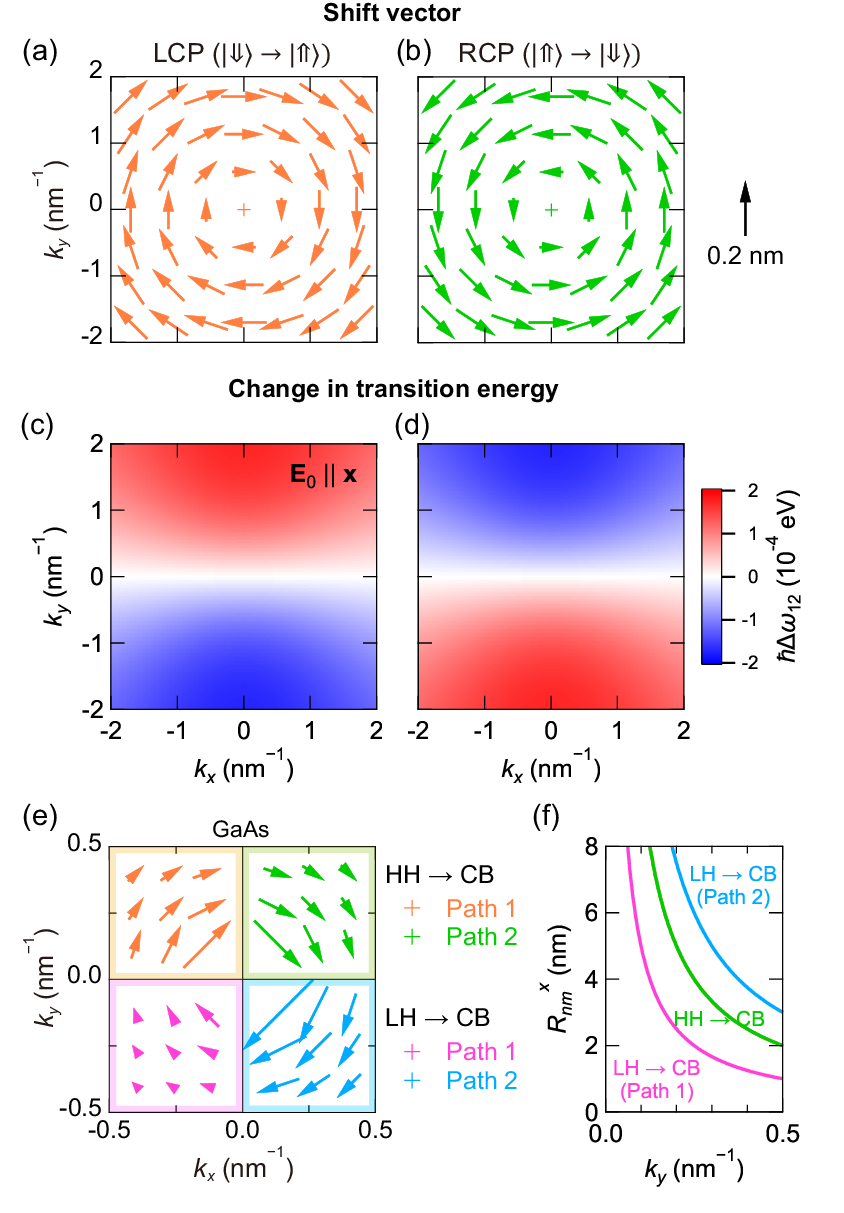}
\caption{(a), (b) Shift vectors for left- and right-circularly polarized light, respectively, in a massive Dirac electron system.
The arrow on the right side corresponds to a length of 0.2 nm.
(c), (d) Magnitude of the transition energy shift for paths 1 and 2 caused by the bias field $E_0 = 10$ kV/cm along the $x$ axis.
(e) Shift vectors for left-circularly polarized light in GaAs.
The first and second quadrants resolve the two transition paths from HH to CB, and
the third and fourth quadrants resolve those from LH to CB.
(f) The $x$ component of the shift vectors at $\kk = (0,k_y,0)$.
}\label{fig:FI-shift}
\end{figure}

Figures \ref{fig:FI-shift}(a) and \ref{fig:FI-shift}(b) illustrate a map of the shift vectors on the $k_z = 0$ plane derived from the massive Dirac electron model.
Here, we distinguish degenerate states with different angular momenta using pseudospins $|\Uparrow\rangle$ and $|\Downarrow\rangle$.
In a lead halide perovskite, the pseudospin corresponds to the ordinary spin in the $s$-like valence band and the total angular momentum $j_z = \pm1/2$ in the $p$-like conduction band \cite{Volosniev2023}.
Panel (a) illustrates the shift vector for $|\Downarrow\rangle\to|\Uparrow\rangle$ driven by left-circularly polarized light,
and panel (b) illustrates that for $|\Uparrow\rangle\to|\Downarrow\rangle$ driven by right-circularly polarized light.
Figures \ref{fig:FI-shift}(c) and \ref{fig:FI-shift}(d) illustrate the corresponding changes in the transition energy under a bias field in the $x$ direction.
The transition energy increases (decreases) at $k_y > 0$ ($k_y < 0$) for left-circularly polarized light, whereas an inverse behavior is observed for right-circularly polarized light.
This contrast results in a photocurrent $\Jj_{\mathrm{inj,ene}}$ in the $y$ direction, with a polarity depending on the helicity of light.

GaAs exhibits a remarkable behavior.
Figure \ref{fig:FI-shift}(e) shows the shift vectors on the $k_z = 0$ plane calculated for the left-circularly polarized light.
The four quadrants distinguish the four transition paths: two from HH to CB and two from LH to CB, each arising from the two-fold degeneracy of every band.
At each $\kk$ point, all four shift vectors point in the same direction, and are inverted for right-circularly polarized light (not shown). 
Notably, all shift vectors diverge as $1/k$ at $k\to0$, as illustrated in Fig. \ref{fig:FI-shift}(f).
In addition, they do not depend on material parameters, such as the band gap energy and effective mass.
For example, the shift vectors for the transition $\text{HH}\to\text{CB}$ are given by $\Rr_{12}(\ee_{\mathrm{L}}) = -(-k_y,k_x,0)/k(k\mp k_z)$ for $\ee_{\mathrm{L}} = (1/\sqrt{2})(1,\i,0)$.
Insensitiveness to the details of the band structure and appearance of a singularity reflect the topological character of the valence band.
Similar to $\Delta\Ww_{nm}$, the divergence in $\Delta\omega_{nm}(\ee)$ causes resonance in the field-induced CPGE as shown later.

\section{Anomalous velocity of photocarriers}

We briefly mention the remaining correction to Eq. \eqref{eq:Jinj_1},
\begin{align}
\dot{\Jj}_{\mathrm{inj,ano}}& = \frac{e^3}{2\hbar^2}\sum_{\kk}\sum_n\sum_{m>n}f_{nm}(\boldsymbol{\Gamma}_{nm}\cdot\Ww_{nm})\nonumber\\
&\quad\times\left[-\frac{e}{\hbar}\Ee_0\times(\Ww_n-\Ww_m)\right].\label{eq:Jinjano_1}
\end{align}
The second line represents the difference between the anomalous velocities of the initial and final states, 
which is considered as a correction of $\nabla_{\kk}\omega_{nm}$ in Eq. \eqref{eq:Jinj_1}.
This is the most conventional mechanism of the photovoltaic Hall effect and is referred to differently depending on the degrees of freedom associated with the Berry curvature $\Ww_n$.
The most familiar example is the inverse spin Hall effect, where optically excited spin-polarized carriers generate a transverse current under a bias field \cite{Bakun1984,Miah2007,Yin2011,Yu2012,Okamoto2014,Virk2011,Priyadarshi2015,Fujimoto2024}.
The valley Hall effect, which arises from the valley degree of freedom in two-dimensional transition metal dichalcogenides, is also well known \cite{Xiao2012,Mak2014}.
Other examples include the orbital Hall effect, associated with the orbital angular momentum \cite{Bernevig2005,Go2021}, and the isospin Hall effect, which is specific to Dirac semimetals \cite{Murotani2024}.
To be precise, Eq. \eqref{eq:Jinjano_1} describes only the intrinsic part of these effects;
extrinsic processes induced by scatterers, such as side jump and skew scattering \cite{Sinova2015}, are not included.
Although they are essential in the low-frequency regime, they can be neglected on ultrafast timescales before the occurrence of scattering events \cite{Fujimoto2024}.
In GaAs, $\Ww_n = Q\kk/2k^3$ $(n = 2,3)$ exhibits a singularity at $\kk = 0$ because of the aforementioned monopole charges.
This causes a resonant enhancement of $\Jj_{\mathrm{inj,ano}}$ at the bandgap energy, as discussed next.

\section{Effect of degeneracy points}

\begin{figure}[t]
\centering
\includegraphics[width=\columnwidth]{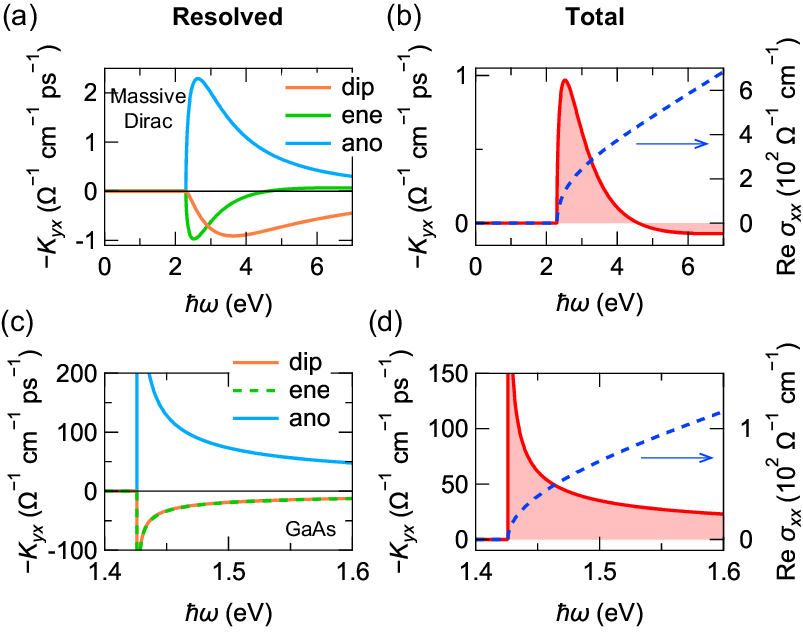}
\caption{(a) Three contributions to the field-induced CPGE in a massive Dirac electron system.
The electric field amplitude of light is set to 100 kV/cm.
(b) The sum of the three contributions in panel (a).
The dashed line represents the real part of optical conductivity in equilibrium.
(c) Three contributions to the field-induced CPGE in GaAs.
$K_{\mathrm{dip}}^{yx}$ (orange) and $K_{\mathrm{ene}}^{yx}$ (green) almost overlap.
The electric field amplitude of light is set to 100 kV/cm.
(d) The sum of the three contributions in panel (c).
The dashed line represents the real part of optical conductivity in equilibrium.
}\label{fig:FI-CPGE}
\end{figure}

We condense the above results into a tensor $K_i^{ab}$ such that
\begin{align}
\dot{J}_{\mathrm{inj},i}^a&=\sum_bK_i^{ab}E_0^b,\quad(i=\text{dip, ene, ano})\label{eq:K_1}
\end{align}
where $a$ and $b$ represent the Cartesian coordinates.
Adding the phenomenological damping term $-J_{\mathrm{inj},i}^a/\tau_i$ to the right-hand side of Eq. \eqref{eq:K_1} yields
a DC optical conductivity tensor $\sigma_i^{ab} = \tau_iK_i^{ab}$.
Figure \ref{fig:FI-CPGE}(a) illustrates the dependence of $K_i^{yx}$ on the photon energy of light for the massive Dirac model irradiated by left-circularly polarized light.
Figure \ref{fig:FI-CPGE}(b) shows the sum $K^{yx} = \sum_iK_i^{yx}$ and absorption spectrum.
Our calculations reproduce the sign change in $K^{yx}$ at $\hbar\omega = 4\Delta$ predicted by Ref. \cite{Ahn2022}.
The dominant origin shifts from an anomalous velocity to a field-induced Berry curvature with this sign change.

In comparison, Fig. \ref{fig:FI-CPGE}(c) illustrates $K_i^{yx}$ for GaAs.
All the contributions exhibit singular behavior in the form of $(\hbar\omega-\Eg)^{-1/2}$ around the band gap energy ($\Eg = 1.426$ eV). 
This behavior originates from the above-mentioned topological nature of the valence band.
Although the anomalous velocity of photoexcited carriers dominates the simple sum $K^{yx}$ [Fig. \ref{fig:FI-CPGE}(d)], 
use of a terahertz pulse $\Ee_0(t)$ instead of a static field $\Ee_0$ allows the different mechanisms to be detected separately.
$\Delta\Ww_{nm}$ and $\Delta\omega_{nm}(\ee)$ should follow the slow temporal variation of $\Ee_0(t)$, leading to $\dot{J}_{\mathrm{inj},i}(t)\sim E_0(t)I_1(t)$ for $i = \text{dip}$ and ene,
where $I_1(t)$ denotes the time-dependent light intensity.
By contrast, $\Jj_{\mathrm{inj,ano}}$ is proportional to the anomalous velocity multiplied by the number of carriers; thus, $J_{\mathrm{inj,ano}}(t)\sim E_0(t)\int_{-\infty}^t\d t'~I_1(t')$.
Such a qualitative difference in the dynamic response can be used to distinguish the different mechanisms, as demonstrated by our recent experiment \cite{Fujimoto2025}.
For strong excitations, $\Jj_{\mathrm{inj,dip}}+\Jj_{\mathrm{inj,ene}}$ even became dominant because of the high suppression of $\Jj_{\mathrm{inj,ano}}$ \cite{Fujimoto2025}.
Strikingly, the resonance peak in $\Jj_{\mathrm{inj,dip}}+\Jj_{\mathrm{inj,ene}}$ shifted to a slightly higher energy because of the resonance between the terahertz pulse and HH--LH transitions \cite{Fujimoto2025}.
Although the essence of the field-induced CPGE can be captured by the present theory, such a dynamic effect deserves further investigation.

\section{Dressed state contribution}

Now, we look into the contribution by light-dressed states formed in metallic systems.
For a non-degenerate case, the anomalous Hall current carried by the Fermi surface turns out to be
\begin{align}
\Jj_{\mathrm{dre,ano}}&=e\sum_{\kk}\sum_nf_n\left(-\frac{e}{\hbar}\Ee_0\times\delta\Ww_n\right),\label{eq:Jdreano_3}
\end{align}
with $\delta\Ww_n=\nabla_{\kk}\times\delta\xxi_{nn}$.
As shown by Eq. \eqref{eq:dxi_1} in Appendix \ref{sec:J2}, $\delta\xxi_{nn}$ is proportional to the intensity of light.
Importantly, $\delta\Ww_n$ coincides with the light-induced Berry curvature predicted by Floquet formalism in the lowest order.
Thus, our theory successfully describes the AHE by Floquet states within the third-order nonlinearity. 
It can also be shown that the second-order electric polarization responsible for optical rectification,
\begin{align}
\Pp_{\mathrm{rec,0}}&=e\sum_{\kk}\sum_nf_n\delta\xxi_{nn},\label{eq:POR_2}
\end{align}
is governed by $\delta\xxi_{nn}$.
Given the established role of Berry connection in modern theory of electric polarization \cite{Resta2000,Xiao2010,Vanderbilt2018}, $\delta\xxi_{nn}$ can be interpreted as a light-induced Berry connection, in accord with the relationship $\delta\Ww_n=\nabla_{\kk}\times\delta\xxi_{nn}$.
This work provides the first analytic formulation of the field-induced CPGE, light-induced AHE by Floquet states, and optical rectification using a single theoretical framework.

\section{Comparison with previous studies}

Finally, we compare our results with those in previous studies on the photovoltaic Hall effect.
A general theory of the bulk photovoltaic effect under a bias field has been presented in the literature \cite{Fregoso2019}.
Although its results for the field-induced CPGE were consistent with ours, the physical meaning of each term was not as clear as revealed here.
For example, $\Jj_{\mathrm{inj,dip}}$ in our theory was split into two terms and was interpreted differently from the field-induced Berry curvature in Ref. \cite{Fregoso2019}.
$\Jj_{\mathrm{inj,ene}}$ was also given an interpretation different from the field-induced energy shift.
Our theory provides an alternative perspective for understanding the photovoltaic Hall effect induced by resonant light.
A more recent theory \cite{Ahn2022} rearranged the result in Ref. \cite{Fregoso2019} in terms of Riemannian geometry.
Although this viewpoint reveals an interesting aspect of light--matter interaction, the obtained equations cannot easily be understood from a physical point of view.
For example, the contribution of the Hermitian curvature, which was the main focus of Ref. \cite{Ahn2022}, was a mixture of $\Jj_{\mathrm{inj,ano}}$ and $\Jj_{\mathrm{inj,ene}}$ in our theory.
These two components should be distinguished both for physical interpretation and for practical use,
because the relaxation times for photocurrent generally differ from each other;
$\Jj_{\mathrm{inj,dip}}+\Jj_{\mathrm{inj,ene}}$ decays through momentum relaxation, whereas $\Jj_{\mathrm{inj,ano}}$ decays through the relaxation of the spin or other quantities {that break the} time-reversal symmetry.
To incorporate this qualitative difference, $\Jj_{\mathrm{inj,dip}}+\Jj_{\mathrm{inj,ene}}$ and $\Jj_{\mathrm{inj,ano}}$ should be distinguished.

\section{Conclusion}

In summary, we developed a comprehensive theory of the photovoltaic Hall effect in nonmagnetic materials.
We revealed three contributions in field-induced CPGE: (i) field-induced Berry curvature, (ii) field-induced energy shift, and (iii) anomalous velocity of photoexited carriers. 
These contributions were related to either the Berry curvature or shift vector, demonstrating the geometric aspect of the field-induced CPGE.
The topological character of the valence band in GaAs generated a resonance peak for all three mechanisms.
We also desribed AHE by light-dressed states within the same theoretical framework and revealed its close relationship with optical rectification.
Our results provide a solid foundation for interpreting the experimental results on the photovoltaic Hall effect and a possible counterpart to the Floquet \cite{Oka2009} and strain \cite{Dong2023} engineering of the Berry curvature. 
Extension to correlated systems is an interesting future direction \cite{Parafilo2023}.

\begin{acknowledgments}
This study was supported by JSPS KAKENHI (Grants No. JP24K16988 and No. JP24K00550), JST FOREST (Grant No. JPMJFR2240), and JST CREST (Grant No. JPMJCR20R4).

Y.M. and R.M. conceived the project. 
Y.M. developed the theoretical formalism with help of T.F. and performed numerical calculation{s}.
All authors discussed the result. 
Y.M. wrote the manuscript with feedback from all coauthors.
\end{acknowledgments}

\appendix

\section{Formulation}\label{sec:formulation}

In this appendix, we outline the derivation of equations presented in the main text.

\subsection{Perturbation by a static electric field}\label{sec:framework}

We start from a Schr\"odinger equation expressed in the length gauge \cite{Aversa1995}, 
\begin{align}
\i\hbar\frac{\partial}{\partial t}|\Phi(\kk)\rangle&=[H(\kk)-\i e\Ee\cdot\nabla_{\kk}]|\Phi(\kk)\rangle,\label{eq:Sch_2}
\end{align}
where $\Ee$ is the electric field.
Our goal is to calculate the current density,
\begin{align}
\Jj&=e\sum_{\kk}\langle\Phi(\kk)|\left[\frac{1}{\hbar}\nabla_{\kk}H(\kk)\right]|\Phi(\kk)\rangle.\label{eq:J_1}
\end{align}
For brevity, we omit $\kk$ dependence of every quantity hereafter.

We diagonalize $H$ so that each group of degenerate states forms a block:
\begin{align}
\mathcal{H}&\equiv U^\dag HU
=\operatorname{diag}\left(\epsilon_1I_1,\epsilon_2I_2,\cdots\right),\label{eq:Heq}
\end{align}
where $\epsilon_\nu$ denotes the dispersion relation of the $\nu$-th band, and $I_\nu$ a unit matrix with a dimension equal to its degeneracy.
We label each energy band with a Greek letter ($\nu$, $\mu$, etc.) and each basis in it with a Roman letter ($n$, $m$, etc.).
In this notation, $U_{nm}=|u_m\rangle_n$ can be regarded as the $n$-th component of the $m$-th eigenstate, $|u_m\rangle$.
Below, the energy eigenvalues are arranged in descending order, i.e., $\epsilon_\nu>\epsilon_{\nu+1}$.
Ability to treat degenerate bands is essential because the photovoltaic Hall effect can occur even in nonmagnetic and centrosymmetric materials.
Using a new wavefunction $|\phi\rangle\equiv U^\dag|\Phi\rangle$, Eq. \eqref{eq:Sch_2} is rewritten as
\begin{align}
\i\hbar\frac{\partial}{\partial t}|\phi\rangle&=[\mathcal{H}-e\Ee\cdot(\xxi+\i\nabla_{\kk})]|\phi\rangle,\label{eq:Sch_2.5}
\end{align}
where $\xxi\equiv U^\dag\i\nabla_{\kk}U$ is a Berry connection matrix.

It is apparent that $\xxi+\i\nabla_{\kk}$ serves as a position operator acting on $|\phi\rangle$ \cite{Blount1962,Aversa1995,Sipe2000}.
In particular, its off-diagonal component $\xxi_{\nu\mu}~(\nu\neq\mu)$ acts as a transition dipole moment, which determines the transition matrix element
$\mathcal{M}_{\nu\mu}(\ee)\equiv(\ee\cdot\xxi_{\nu\mu})(\ee^*\cdot\xxi_{\mu\nu})$ for the transition $\mu\to\nu$.
Even in systems with degenerate bands, a relationship between Berry curvature and circular dichroism, similar to Eq. \eqref{eq:CD_2}, holds:
\begin{align}
\Omega_{\nu\mu}^z&=\mathcal{M}_{\nu\mu}(\ee_{\mathrm{R}})-\mathcal{M}_{\nu\mu}(\ee_{\mathrm{L}}),\label{eq:CD_1}
\end{align}
where the band-resolved Berry curvature is defined by $\Ww_{\nu\mu}\equiv\i\xxi_{\nu\mu}\times\xxi_{\mu\nu}$.
Summation of $\Ww_{\nu\mu}$ over $\mu$ gives the total Berry curvature for the $\nu$-th band [see Eq. \eqref{eq:W_1}].

Now we apply a static electric field $\Ee_0$ to the system.
In this case, it is helpful to introduce another wavefunction $|\tilde{\phi}\rangle\equiv\tilde{U}^\dag|\Phi\rangle$,
which transforms Eq. \eqref{eq:Sch_2} into
\begin{align}
\i\hbar\frac{\partial}{\partial t}|\tilde{\phi}\rangle&=(\tilde{\mathcal{H}}-\i e\Ee_0\cdot\nabla_{\kk})|\tilde{\phi}\rangle,\label{eq:Sch_3}
\end{align}
with a modified Hamiltonian $\tilde{\mathcal{H}}\equiv\tilde{U}^\dag H\tilde{U}-e\Ee_0\cdot\tilde{\xxi}$ 
and a modified Berry connection matrix $\tilde{\xxi}\equiv\tilde{U}^\dag\i\nabla_{\kk}\tilde{U}$.
We require $\tilde{\mathcal{H}}$ to be block-diagonal like Eq. \eqref{eq:Heq}.
Up to the first order in $\Ee_0$, this can be done by putting $\tilde{U}=U+\Delta U$ with $(U^\dag\Delta U)_{\nu\mu}=e\Ee_0\cdot\xxi_{\nu\mu}/\hbar\omega_{\nu\mu}$ $(\nu\neq\mu)$,
where $\omega_{\nu\mu}\equiv(\epsilon_\nu-\epsilon_\mu)/\hbar$ is the transition frequency.
We can safely put $(U^\dag\Delta U)_{\nu\nu}=0$.
In the end, $\tilde{\mathcal{H}}$ retains only the block-diagonal components, $\tilde{\mathcal{H}}_{\nu\nu}=\epsilon_\nu I_\nu-e\Ee_0\cdot\xxi_{\nu\nu}$.

\subsection{Correction to transition dipole moment}\label{sec:dipole}

The static field $\Ee_0$ modifies the Berry connection.
The correction in $\tilde{\xxi}=\xxi+\Delta\xxi$ can be expressed as $\Delta\xxi=C\Ee_0$,
with the Berry connection polarizability
\begin{align}
C_{\nu\nu}^{ab}&=-\frac{e}{\hbar}\sum_{\mu\neq\nu}\frac{\xi_{\nu\mu}^a\xi_{\mu\nu}^b+\xi_{\nu\mu}^b\xi_{\mu\nu}^a}{\omega_{\nu\mu}},\label{eq:Cnn}\\
C_{\nu\mu}^{ab}&=\frac{\i e}{\hbar}D^a\left(\frac{\xi_{\nu\mu}^b}{\omega_{\nu\mu}}\right)+\frac{e}{\hbar}\sum_{\lambda\neq\nu,\mu}\left(\frac{\xi_{\nu\lambda}^a\xi_{\lambda\mu}^b}{\omega_{\lambda\mu}}-\frac{\xi_{\nu\lambda}^b\xi_{\lambda\mu}^a}{\omega_{\nu\lambda}}\right).\nonumber\\
&\quad(\nu\neq\mu)\label{eq:Cnm}
\end{align}
Here, $\Dd_{\kk}=(D^x,D^y,D^z)$ represents a generalized \cite{Aversa1995} or covariant \cite{Fregoso2018} derivative defined by 
$\Dd_{\kk}O\equiv-\i[\rr_{\mathrm{i}},O]$ using the intraband position operator $\rr_{\mathrm{i},\nu\mu}\equiv\delta_{\nu\mu}(\xxi_{\nu\nu}+\i\nabla_{\kk})$.
More specifically, $\Dd_{\kk}O_{\nu\mu}=\nabla_{\kk}O_{\nu\mu}-\i(\xxi_{\nu\nu}O_{\nu\mu}-O_{\nu\mu}\xxi_{\mu\mu})$.
Berry curvature is also changed into 
$\tilde{\Ww}_{\nu\mu}\equiv\i\tilde{\xxi}_{\nu\mu}\times\tilde{\xxi}_{\mu\nu}$,
which includes a correction $\Delta\Ww_{\nu\mu}=\Pi_{\nu\mu}\Ee_0$.
The band-resolved Berry curvature polarizability $\Pi_{\nu\mu}$ is given by
\begin{align}
\Pi_{\nu\mu}^{ad}&\equiv\i\sum_{bc}\epsilon^{abc}\mathcal{T}_{\nu\mu}^{bcd},
\end{align}
where
\begin{align}
\mathcal{T}_{\nu\mu}^{abc}&\equiv\xi_{\nu\mu}^aC_{\mu\nu}^{bc}+C_{\nu\mu}^{ac}\xi_{\mu\nu}^b.\label{eq:T_1}
\end{align}

\subsection{Correction to transition energy}\label{sec:energy}

Next, we examine the velocity operator 
$\tilde{\vv}\equiv\tilde{U}^\dag(\hbar^{-1}\nabla_{\kk}H)\tilde{U}$ under the electric field.
The off-diagonal components are found to be
\begin{align}
\tilde{\vv}_{\nu\mu}&=\i\omega_{\nu\mu}\tilde{\xxi}_{\nu\mu}+\frac{e}{\hbar}(\Ee_0\cdot\Dd_{\kk})\xxi_{\nu\mu}.\quad(\nu\neq\mu)\label{eq:vnm_1}
\end{align}
At $\Ee_0=0$, Eq. \eqref{eq:vnm_1} recovers a well-known relationship between position and velocity operators, 
$\vv_{\nu\mu}=\i\omega_{\nu\mu}\xxi_{\nu\mu}$ for $\nu\neq\mu$.
This equation lets us expect that Eq. \eqref{eq:vnm_1} should include corrections of $\xxi_{\nu\mu}$ and $\omega_{\nu\mu}$ induced by $\Ee_0$.
The former clearly appears in the first term on the right-hand side of Eq. \eqref{eq:vnm_1}.
To demonstrate the latter effect, we consider a non-degenerate case.
Multiplied by the polarization vector $\ee$, Eq. \eqref{eq:vnm_1} gives
\begin{align}
\ee\cdot\tilde{\vv}_{nm}&=\i[\omega_{nm}+\Delta\omega_{nm}(\ee)+(\text{imag.})](\ee\cdot\tilde{\xxi}_{nm}),\label{eq:vnm_2}
\end{align}
with
\begin{align}
\hbar\Delta\omega_{nm}(\ee)&\equiv-e\Ee_0\cdot\Rr_{nm}(\ee).\label{eq:Dwnm_1}
\end{align}
Here, $\Rr_{nm}(\ee)\equiv\xxi_{nn}-\xxi_{mm}-\nabla_{\kk}\operatorname{arg}(\ee\cdot\xxi_{nm})$ 
is the well-known shift vector, which measures the spatial movement of an electron cloud accompanying an interband transition $m\to n$ \cite{Sipe2000}.
Given such a spatial shift, the electrostatic potential for the initial and final states should differ by $\hbar\Delta\omega_{nm}(\ee)$ under $\Ee_0$.
Equation \eqref{eq:vnm_2} tells that such a modification in the transition energy is naturally encoded in Eq. \eqref{eq:vnm_1}, including polarization dependence of the shift vector.
Even for systems with degenerate bands, appropriate choice of bases resolves the mutually independent pairs of inital and final states, making the above argument valid (see Appendix \ref{sec:SVD}).
We note that (imag.) in Eq. \eqref{eq:vnm_2}, indicating an imaginary number, is not important because it does not affect the transition probability.

\subsection{Anomalous velocity and acceleration}

The other effects of $\Ee_0$ are well-known.
First, the diagonal component of the modified velocity operator takes the form of 
\begin{align}
\tilde{\vv}_{\nu\nu}&=\frac{1}{\hbar}\nabla_{\kk}\epsilon_\nu-\frac{e}{\hbar}\Ee_0\times\Ww_\nu,\label{eq:vnn_1}
\end{align}
where
\begin{align}
\Ww_\nu&\equiv\nabla_{\kk}\times\xxi_{\nu\nu}-\i\xxi_{\nu\nu}\times\xxi_{\nu\nu}=\sum_{\mu\neq\nu}\Ww_{\nu\mu},\label{eq:W_1}
\end{align}
is the Berry curvature of the $\nu$-th band.
The first and second terms in Eq. \eqref{eq:vnn_1} correspond to the group and anomalous velocities, respectively \cite{Nagaosa2010,Xiao2010}.

The last effect of $\Ee_0$ is acceleration.
To see this, we complete the formalism by adding an optical field $\Ee_1$ to the total electric field, $\Ee=\Ee_0+\Ee_1$.
Equation \eqref{eq:Sch_2} is then transformed into
\begin{align}
\i\hbar\frac{\partial}{\partial t}|\tilde{\phi}\rangle&=(\tilde{\mathcal{H}}-e\Ee_1\cdot\tilde{\xxi}-\i e\Ee\cdot\nabla_{\kk})|\tilde{\phi}\rangle.\label{eq:Sch_4}
\end{align}
For a density matrix, $\rho\equiv|\tilde{\phi}\rangle\langle\tilde{\phi}|$, the equation of motion is given by
\begin{align}
\dot{\rho}&=-\frac{\i}{\hbar}[\tilde{\mathcal{H}}-e\Ee_1\cdot\tilde{\xxi},\rho]-\frac{e}{\hbar}(\Ee\cdot\nabla_{\kk})\rho.\label{eq:EoM_1}
\end{align}
Compared to the case of $\Ee_0=0$, Eq. \eqref{eq:EoM_1} includes three changes: 
(i) correction to energy, $\mathcal{H}\to\tilde{\mathcal{H}}$, 
(ii) correction to transition dipole moment, $\xxi\to\tilde{\xxi}$, and 
(iii) an additional term, $-(e/\hbar)(\Ee_0\cdot\nabla_{\kk})\rho$. 
The last one is responsible for acceleration of electrons.
Equation \eqref{eq:EoM_1} can be extended to a time-dependent $\Ee_0$ as long as its frequency lies far below the interband transition frequencies.

We solve Eq. \eqref{eq:EoM_1} neglecting scattering processes.
This is done by putting a time-dependent initial condition,
\begin{align}
\rho_{\nu\nu}^{(0)}&=f_\nu+\frac{e}{\hbar}(\Aa_0\cdot\nabla_{\kk})f_\nu,\label{eq:rhonn0_1}
\end{align}
before the arrival of the optical field $\Ee_1$.
Here, the superscript in parentheses indicates the order with respect to $\Ee_1$, and $f_\nu$ the Fermi-Dirac distribution function for energy $\epsilon_\nu$.
In reality, scattering processes resist the acceleration by $\Ee_0$, which should generate a constant distribution function instead of Eq. \eqref{eq:rhonn0_1}.
Scattering processes can also take part in photocurrent generation by themselves \cite{Sturman2020,Dai2021,Zhu2024}.
However, it is challenging to incorporate these effects in a comprehensive manner, since there exist complex scattering channels including side jump \cite{Berger1970} and skew scattering \cite{Smit1958} known in anomalous and spin Hall effects \cite{Nagaosa2010,Sinova2015}.
Rather than elaborating a model that takes into account these complexities, we aim at obtaining a clear physical picture of photovoltaic Hall effect in a simplified situation, 
which helps comparison with Floquet theory and lays the foundation for future extension.
Equation \eqref{eq:rhonn0_1} is indeed justified in a time scale shorter than the scattering times, which is relevant to recent ultrafast experiments using terahertz spectroscopy \cite{Murotani2023,Murotani2024,Fujimoto2024,Fujimoto2025,Yoshikawa2022,Hirai2023}.

\section{Calculation of photocurrent}\label{sec:J2}

\subsection{Injection current}\label{sec:injection}

Solving Eq. \eqref{eq:EoM_1} to obtain the nonlinear current requires a lengthy calculation.
Here, we only present the final results.
The second-order nonlinear current in the presence of a static field is classified into 
\begin{align}
\Jj^{(2)}&=\Jj_{\mathrm{jer}}+\Jj_{\mathrm{inj}}+\Jj_{\mathrm{shi}}+\dot{\Pp}_{\mathrm{rec}}+\Jj_{\mathrm{dre}}.\label{eq:J2_1}
\end{align}
The first term, called jerk current, arises from acceleration of photoexcited carriers \cite{Fregoso2018,Ventura2021,Fregoso2021}.
The second and third, i.e., injection and shift currents, originate from momentum asymmetry and spatial transfer of photoexcited carriers, respectively \cite{Sipe2000}.
The fourth is the well-known optical rectification, and the last originates from response on the Fermi surface.
For monochromatic illumination, $\ddot{\Jj}_{\mathrm{jer}}$, $\dot{\Jj}_{\mathrm{inj}}$, $\Jj_{\mathrm{shi}}$, and $\Pp_{\mathrm{rec}}$ become constant \cite{Fregoso2019}, while $\Jj_{\mathrm{dre}}$ includes terms with different dependencies.
In a nonmagnetic material, only $\Jj_{\mathrm{inj}}$ and $\Jj_{\mathrm{dre}}$ yield the photovoltaic Hall effect.
We therefore omit the other terms.

Injection current in the presence of a static field is decomposed into
\begin{align}
\Jj_{\mathrm{inj}}&=\Jj_{\mathrm{inj,0}}+\Jj_{\mathrm{inj,dip}}+\Jj_{\mathrm{inj,ene}}+\Jj_{\mathrm{inj,ano}}+\Jj_{\mathrm{inj,fer}}.\label{eq:Jinj_full}
\end{align}
The first term corresponds to the usual injection current in the absence of $\Ee_0$, which follows
\begin{align}
\dot{\Jj}_{\mathrm{inj,0}}&=-\frac{e^3}{\hbar^2}\sum_{\kk}\sum_\nu\sum_{\mu>\nu}f_{\nu\mu}(\nabla_{\kk}\omega_{\nu\mu})\sum_{bc}\Gamma_{\nu\mu}^{bc}M_{\nu\mu}^{bc},\label{eq:Jinj_0}
\end{align}
where $f_{\nu\mu}\equiv f_\nu-f_\mu$ and $M_{\nu\mu}^{ab}\equiv\Tr\xi_{\nu\mu}^a\xi_{\mu\nu}^b$.
$\Gamma_{\nu\mu}^{bc}$ is defined as $\Gamma_{\nu\mu}^{bc}\equiv(1/2)(\partial^3G_{\nu\mu}^bG_{\mu\nu}^c/\partial t^3)$ using 
\begin{align}
\mathbf{F}_{\nu\mu}&\equiv\int_{-\infty}^t\d t'~\e^{\i\omega_{\nu\mu}t'}\Ee_1(t'),\ 
\mathbf{G}_{\nu\mu}\equiv\int_{-\infty}^t\d t'~\mathbf{F}_{\nu\mu}(t'),\label{eq:GG_1}
\end{align}
and is equal to
\begin{align}
\Gamma_{\nu\mu}^{bc}&=2\pi E_1^bE_1^{c*}\delta(\omega-\omega_{\nu\mu})+(\nu\leftrightarrow\mu,b\leftrightarrow c),\label{eq:Gamma_2}
\end{align}
for a monochromatic wave, $\Ee_1(t)=\Ee_1\e^{-\i\omega t}+\Ee_1^*\e^{\i\omega t}$.
In the presence of time-reversal symmetry, $\Jj_{\mathrm{inj,0}}$ can be expressed alternatively as
\begin{align}
\dot{\Jj}_{\mathrm{inj,0}}&=\frac{e^3}{2\hbar^2}\sum_{\kk}\sum_\nu\sum_{\mu>\nu}f_{\nu\mu}(\nabla_{\kk}\omega_{\nu\mu})(\boldsymbol{\Gamma}_{\nu\mu}\cdot\Ww_{\nu\mu}^{\mathrm{tr}}),\label{eq:Jinj_2}
\end{align}
where $\Ww_{\nu\mu}^{\mathrm{tr}}\equiv\Tr\Ww_{\nu\mu}$ and $\Gamma_{\nu\mu}^a\equiv\i\sum_{bc}\epsilon^{abc}\Gamma_{\nu\mu}^{bc}$.
For a monochromatic wave, we have
$\boldsymbol{\Gamma}_{\nu\mu}=2\pi\i(\Ee_1\times\Ee_1^*)\delta(\omega-\omega_{\nu\mu})$,
which is parallel to the propagation direction of light and is proportional to the degree of circular polarization.
Equation \eqref{eq:Jinj_1} in the main text derives from \eqref{eq:Jinj_2}.

The second term in Eq. \eqref{eq:Jinj_full} is defined by
\begin{align}
\dot{\Jj}_{\mathrm{inj,dip}}&=-\frac{e^3}{\hbar^2}\sum_{\kk}\sum_\nu\sum_{\mu>\nu}f_{\nu\mu}(\nabla_{\kk}\omega_{\nu\mu})\sum_{bcd}\Gamma_{\nu\mu}^{bc}T_{\nu\mu}^{bcd}E_0^d,\label{eq:Jinjdip_0}
\end{align}
where $T_{\nu\mu}^{bcd}\equiv\Tr\mathcal{T}_{\nu\mu}^{bcd}$.
This contribution arises from the correction in transition dipole moment discussed in Sec. \ref{sec:dipole}.
In the presence of time-reversal symmetry, we obtain an alternative expression,
\begin{align}
\dot{\Jj}_{\mathrm{inj,dip}}&=\frac{e^3}{2\hbar^2}\sum_{\kk}\sum_\nu\sum_{\mu>\nu}f_{\nu\mu}(\nabla_{\kk}\omega_{\nu\mu})(\boldsymbol{\Gamma}_{\nu\mu}\cdot\Delta\Ww_{\nu\mu}^{\mathrm{tr}}),\label{eq:Jinjdip_2}
\end{align}
where $\Delta\Ww_{\nu\mu}^{\mathrm{tr}}\equiv\Tr\Delta\Ww_{\nu\mu}$.
Equation \eqref{eq:Jinjdip_2} unifies the third and fourth terms in Eq. (137) of Ref. \cite{Fregoso2019} in a physically clear form.
Equation \eqref{eq:Jinjdip_1} in the main text derives from \eqref{eq:Jinjdip_2}.

The third term in Eq. \eqref{eq:Jinj_full} is defined by
\begin{align}
\dot{\Jj}_{\mathrm{inj,ene}}&=\frac{e^4}{\hbar^3}\sum_{\kk}\sum_\nu\sum_{\mu>\nu}f_{\nu\mu}\sum_{bcd}(\nabla_{\kk}\Gamma_{\nu\mu}^{bc})S_{\nu\mu}^{bcd}E_0^d,\label{eq:Jinjene_0}
\end{align}
where $S_{\nu\mu}^{bcd}\equiv\Tr\mathcal{S}_{\nu\mu}^{bcd}$ and 
\begin{align}
\mathcal{S}_{\nu\mu}^{abc}&\equiv-\frac{\i}{2}[\xi_{\nu\mu}^a(D^c\xi_{\mu\nu}^b)-(D^c\xi_{\nu\mu}^a)\xi_{\mu\nu}^b].\label{eq:S_1}
\end{align}
An identity $\nabla_{\kk}\Gamma_{\nu\mu}^{bc}=(\partial\Gamma_{\nu\mu}^{bc}/\partial\omega_{\nu\mu})\nabla_{\kk}\omega_{\nu\mu}$ suggests that $\Jj_{\mathrm{inj,ene}}$ should be related to the field-induced energy shift in the interband transitions.
To confirm this, here we consider a non-degenerate case by replacing $(\nu,\mu)\to(n,m)$.
Since $\Gamma_{nm}^{bc}$ vanishes for nonresonant light, we can adopt rotating wave approximation, which allows us to put $\Gamma_{nm}^{bc}=\Gamma_{nm}e^be^{c*}$.
After some of algebra, we obtain
\begin{align}
\dot{\Jj}_{\mathrm{inj,ene}}&=-\frac{e^3}{\hbar^2}\sum_{\kk}\sum_n\sum_{m>n}f_{nm}(\nabla_{\kk}\omega_{nm})\nonumber\\
&\quad\times\sum_{bc}\Delta\omega_{nm}(\ee)\frac{\partial\Gamma_{nm}^{bc}}{\partial\omega_{nm}}M_{nm}^{bc}.\label{eq:Jinjene_2}
\end{align}
As expected, Eq. \eqref{eq:Jinjene_2} can be interpreted as a correction to Eq. \eqref{eq:Jinj_0} by the energy shift $\hbar\Delta\omega_{nm}(\ee)$ determined by the shift vector [Eq. \eqref{eq:Dwnm_1}].
A similar role of the shift vector has been recognized in the context of Landau-Zener tunneling \cite{Kitamura2020,Takayoshi2021,Morimoto2023}.
It is also quite analogous to the anomalous distribution mechanism of the AHE in magnets, where Fermi's golden rule for impurity scattering is modified by the potential energy difference added by side jump under a bias electric field \cite{Sinitsyn2007,Sinitsyn2008}.
Even for degenerate bands, singular value decomposition resolves the shift vector $\Rr_{\nu\mu,n}(\ee)$ and the energy change $\hbar\Delta\omega_{\nu\mu,n}(\ee)$ for each path $n$, reducing Eq. \eqref{eq:Jinjene_1} to a form equivalent to Eq. \eqref{eq:Jinjene_2} (see Appendix \ref{sec:SVD}).
In nonmagnetic materials, we can obtain an alternative expression:
\begin{align}
\dot{\Jj}_{\mathrm{inj,ene}}&=\frac{e^3}{2\hbar^2}\sum_{\kk}\sum_\nu\sum_{\mu>\nu}f_{\nu\mu}\sum_b(\nabla_{\kk}\Gamma_{\nu\mu}^b)W_{\nu\mu}^b,\label{eq:Jinjene_3}
\end{align}
where $W_{\nu\mu}^a\equiv-(\i e/\hbar)\sum_{bcd}\epsilon^{abc}S_{\nu\mu}^{bcd}E_0^d$ contains the information of energy shift in a compact form.
$\Jj_{\mathrm{inj,ene}}$ in this form is equivalent to the second term in Eq. (137) of Ref. \cite{Fregoso2019}.
Equation \eqref{eq:Jinjene_1} in the main text follows from \eqref{eq:Jinjene_2}.

We note that both of $\Jj_{\mathrm{inj,dip}}$ and $\Jj_{\mathrm{inj,ene}}$ generally appear simultaneously because they are governed by similar tensors, $\mathcal{T}_{\nu\mu}^{bcd}$ and $\mathcal{S}_{\nu\mu}^{bcd}$.
Actually, a simple proportionality $\mathcal{S}_{\nu\mu}^{bcd}=(\hbar\omega_{\nu\mu}/4e)\mathcal{T}_{\nu\mu}^{bcd}$ holds for a two-band system with a Hamiltonian linear in $\kk$, such as the massive Dirac electron model.
A convenient relationship $\mathbf{W}_{\nu\mu}=-(\omega_{\nu\mu}/4)\Delta\Ww_{\nu\mu}^{\mathrm{tr}}$ holds in such a case.

\begin{figure}[t]
\centering
\includegraphics[width=\columnwidth]{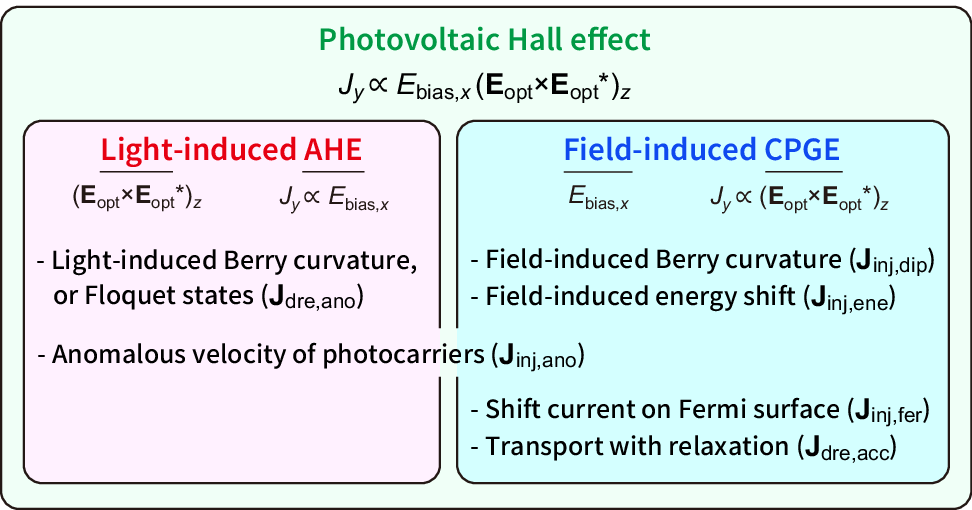}
\caption{
Classification of photovoltaic Hall effect in nonmagnetic materials.
}\label{fig:class}
\end{figure}

The fourth term in Eq. \eqref{eq:Jinj_full} arises from the anomalous velocity of photoexcited electrons and holes.
It obeys
\begin{align}
\dot{\Jj}_{\mathrm{inj,ano}}&=-\frac{e^3}{\hbar^2}\sum_{\kk}\sum_\nu\sum_{\mu>\nu}f_{\nu\mu}\sum_{bc}\Gamma_{\nu\mu}^{bc}\nonumber\\
&\quad\times\Tr\left(\Delta\vv_{\nu\nu}\mathcal{M}_{\nu\mu}^{bc}-\Delta\vv_{\mu\mu}\mathcal{M}_{\mu\nu}^{cb}\right),
\end{align}
where $\Delta\vv_{\nu\nu}\equiv-(e/\hbar)\Ee_0\times\Ww_\nu$ denotes the anomalous velocity, and $\mathcal{M}_{\nu\mu}^{ab}\equiv\xi_{\nu\mu}^a\xi_{\mu\nu}^b$.
Similar to the preceding terms, the rate of current generation $\dot{\Jj}_{\mathrm{inj,ano}}$ becomes a constant for a combination of a static bias field and monochromatic light, which is the reason why it is formally classified into the injection current according to the criteria in Ref. \cite{Fregoso2019}.
However, there is also a qualitative difference between this term and the others.
As discussed above, $\Jj_{\mathrm{inj,dip}}+\Jj_{\mathrm{inj,ene}}$ originates from modification of the transition probability by $\Ee_0$, so that it appears only when the light and bias fields are applied simultaneously.
By contrast, in the case of $\Jj_{\mathrm{inj,ano}}$, $\Ee_0$ does not affect the excitation process but changes the intraband velocity, so that this current can be generated even when the bias field is applied after a light pulse excites the system.
To cover such a persisting response, it is better to recognize $\Jj_{\mathrm{inj,ano}}$ as a light-induced AHE originating from time-reversal symmetry breaking by photoexcited carriers, rather than a field-induced CPGE which can only explain the situation with temporally overlapping light and bias fields.
Nevertheless, the case with a static bias field allows both interpretations, as indicated in Fig. \ref{fig:class}.
In a nonmagnetic system, one can obtain an alternative expression,
\begin{align}
\dot{J}_{\mathrm{inj,ano}}^a&=-\frac{e^4}{2\hbar^3}\sum_{\kk}\sum_\nu\sum_{\mu>\nu}f_{\nu\mu}\sum_{bcd}\epsilon^{abc}E_0^b\nonumber\\
&\quad\times\Tr(\Omega_\nu^c\Omega_{\nu\mu}^d+\Omega_\mu^c\Omega_{\mu\nu}^d)\Gamma_{\nu\mu}^d,\label{eq:Jinjano_2}
\end{align}
which shows dependence on the helicity of light through $\boldsymbol{\Gamma}_{\nu\mu}$.
This form is equivalent to the first term in Eq. (137) of Ref. \cite{Fregoso2019}.
A set of tensor indices $(abcd)=(yxzz)$ matters when light impinges the sample normally and the AHE is detected within the sample plane.
For two-band systems, we have $\Ww_1=\Ww_{12}$ and $\Ww_2=\Ww_{21}$, which further simplifies calculation.
The massive Dirac electron model yields a Berry curvature $\Ww_{1,2}\simeq2\lambda\boldsymbol{\sigma}$ in the neighborhood of band gap.
Since $(\sigma_z)^2$ has a nonzero trace, Eq. \eqref{eq:Jinjano_2} does not vanish.
If the operator $\boldsymbol{\sigma}$ is associated with spin, it can be regarded as a manifestation of inverse spin Hall effect.
In graphene, we obtain $\Jj_{\mathrm{inj,ano}}=0$ due to a vanishing Berry curvature.

The last term in Eq. \eqref{eq:Jinj_full} is qualitatively different from the others.
It is given by
\begin{align}
J_{\mathrm{inj,fer}}^a&=-\frac{e^4}{\hbar^3}\sum_{\kk}\sum_\nu\sum_{\mu>\nu}(\Aa_0\cdot\nabla_{\kk}f_{\nu\mu})\sum_{bc}\Gamma_{\nu\mu}^{bc}S_{\nu\mu}^{bca}.\label{eq:Jinjfer_1}
\end{align}
Physically, this contribution can be interpreted as a shift current generated on the carrier distribution function accelerated by the bias field.
A similar role of acceleration in current-induced second harmonic generation has been discussed in literature \cite{Takasan2021}.
Similar to the other terms, $\Jj_{\mathrm{inj,fer}}$ vanishes for nonresonant light, shows a constant generation rate $\dot{\Jj}_{\mathrm{inj,fer}}$ for monochromatic illumination in the absence of relaxation, and exhibits helicity dependence in a nonmagnetic case.
However, unlike the other terms, it appears only when a Fermi surface exists and is directly involved in interband transitions.
This is a rather rare situation in the context of photovoltaic Hall effect, so we do not delve into $\Jj_{\mathrm{inj,fer}}$ in more detail.

\subsection{Contribution by dressed states}\label{sec:dressed}

The photocurrent carried by light-dressed electrons can be devided into 
\begin{align}
\Jj_{\mathrm{dre}}&=\Jj_{\mathrm{dre,0}}+\Jj_{\mathrm{dre,acc}}+\Jj_{\mathrm{dre,ano}}.\label{eq:Jdre_1}
\end{align}
The first term, 
\begin{align}
\Jj_{\mathrm{dre,0}}&=\frac{e}{\hbar}\sum_{\kk}\sum_\nu\Tr f_\nu\nabla_{\kk}\delta\epsilon_\nu,
\end{align}
with
\begin{align}
\delta\epsilon_\nu&=\frac{e^2}{2\hbar}(\Aa_1\cdot\nabla_{\kk})(\Aa_1\cdot\vv_{\nu\nu})+\frac{e^2}{2\hbar}(\Aa_1\times\Ee_1)\cdot\Ww_\nu\nonumber\\
&\quad+\frac{\i e^2}{2\hbar}\sum_{\mu\neq\nu}\sum_{bc}(F_{\nu\mu}^b\dot{F}_{\mu\nu}^c-\dot{F}_{\nu\mu}^bF_{\mu\nu}^c)\mathcal{M}_{\nu\mu}^{bc},\label{eq:de_1}
\end{align}
is independent of $\Ee_0$ and accounts for the Fermi-surface contribution to the bulk photovoltaic effect \cite{Gao2021}.
It includes the nonlinear Hall effect by Berry curvature dipole \cite{Sodemann2015} and Berry connection polarizability \cite{Gao2014,Liu2021,Liu2022}, as well as the metallic jerk current \cite{Matsyshyn2019,Fregoso2019}.
The second term,
\begin{align}
\Jj_{\mathrm{dre,acc}}&=\frac{e^2}{\hbar^2}\sum_{\kk}\sum_\nu\Tr(\Aa_0\cdot\nabla_{\kk}f_\nu)\nabla_{\kk}\delta\epsilon_\nu,\label{eq:Jdreacc_1}
\end{align}
emerges from acceleration of dressed states by $\Ee_0$.
It may contribute to the photovoltaic Hall effect when combined with energy-dependent scattering \cite{Durnev2021}, so we have included it in Fig. \ref{fig:class}.
However, we do not focus on it because such an extrinsic effect lies outside the scope of this paper.
The last term,
\begin{align}
\Jj_{\mathrm{dre,ano}}&=-\frac{e^2}{\hbar}\sum_{\kk}\sum_\nu\Tr f_\nu\Dd_{\kk}(\Ee_0\cdot\delta\xxi_{\nu\nu})\nonumber\\
&\quad-\frac{e^2}{\hbar}\frac{\partial}{\partial t}\sum_{\kk}\sum_\nu\Tr f_\nu(\Aa_0\cdot\Dd_{\kk})\delta\xxi_{\nu\nu},\label{eq:Jdreano_2}
\end{align}
describes the light-induced AHE caused by the anomalous velocity of light-dressed electrons.
Indeed, for monochromatic illumination that makes the light-induced Berry connection $\delta\xxi_{\nu\nu}$ constant, Eq. \eqref{eq:Jdreano_2} is reduced to 
\begin{align}
\Jj_{\mathrm{dre,ano}}&=e\sum_{\kk}\sum_\nu\Tr f_\nu\left(-\frac{e}{\hbar}\Ee_0\times\delta\Ww_\nu\right),\label{eq:Jdreano_1}
\end{align}
where $\delta\Ww_\nu=\Dd_{\kk}\times\delta\xxi_{\nu\nu}$ denotes the light-induced Berry curvature.
The explicit form of $\delta\xxi_{\nu\nu}$ is given by
\begin{align}
\delta\xi_{\nu\nu}^a
&=-\frac{e^2}{2\hbar^2}(\Aa_1\cdot\Dd_{\kk})(\Aa_1\times\Ww_\nu)^a\nonumber\\
&\quad-\frac{e}{2\hbar}\sum_b(\Aa_1\times\Ee_1)^b\Pi_\nu^{ba}\nonumber\\
&\quad-\frac{\i e}{2\hbar}\sum_{\mu\neq\nu}\sum_{bc}(F_{\nu\mu}^b\dot{F}_{\mu\nu}^c-\dot{F}_{\nu\mu}^bF_{\mu\nu}^c)\mathcal{T}_{\nu\mu}^{bca}\nonumber\\
&\quad+\frac{e^2}{2\hbar^2}\sum_{\mu\neq\nu}\sum_{bc}(G_{\nu\mu}^b\dot{F}_{\mu\nu}^c+\dot{F}_{\nu\mu}^bG_{\mu\nu}^c)\mathcal{S}_{\nu\mu}^{bca},\label{eq:dxi_1}
\end{align}
where
\begin{align}
\Pi_\nu^{ad}&\equiv\sum_{bc}\epsilon^{abc}D^bC_{\nu\nu}^{cd}=\sum_{\mu\neq\nu}\Pi_{\nu\mu}^{ad},\label{eq:Pin_1}
\end{align}
is the Berry curvature polarizability which describes the correction $\Delta\Ww_\nu=\Pi_\nu\Ee_0$ to Berry curvature.
One can show that $\delta\xxi_{\nu\nu}$ also describes the optical rectification,
\begin{align}
\Pp_{\mathrm{rec,0}}&=e\sum_{\kk}\sum_\nu\Tr f_\nu\delta\xxi_{\nu\nu}.\label{eq:POR_1}
\end{align}
Thus, $\delta\xxi_{\nu\nu}$ can be naturally interpreted as spatial shift of electron wavefunctions under a light field.
One can also show that Eq. \eqref{eq:dxi_1} is equivalent to the prediction by the Floquet formalism.

\section{Singular value decomposition}\label{sec:SVD}

\begin{figure}[t]
\centering
\includegraphics[width=0.9\columnwidth]{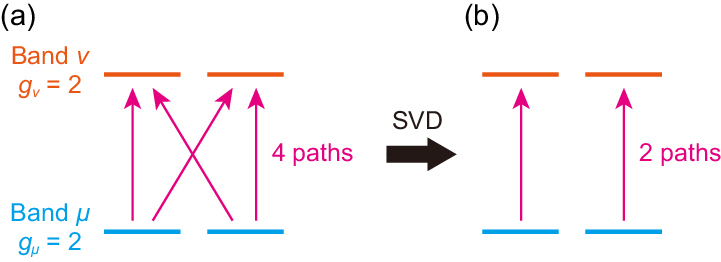}
\caption{(a) Before singular value decomposition (SVD), there are $g_{\nu}\times g_{\mu}$ excitation paths available for the interband transition from band $\mu$ to band $\nu$.
The case of $g_\nu=g_\mu=2$ is shown.
(b) SVD resolves mutually independent pairs of initial and final states, the number of which is less than $\operatorname{min}(g_\nu,g_\mu)$. 
}\label{fig:SVD}
\end{figure}

When the degeneracy of band $\nu$ is written as $g_\nu$, the number of possible excitation paths from band $\mu$ to band $\nu$ amounts to $g_\nu g_\mu$ for an arbitrary choice of bases, as shown in Fig. \ref{fig:SVD}(a).
Singular value decomposition helps us reduce it to less than $\operatorname{min}(g_\nu,g_\mu)$, as shown in Fig. \ref{fig:SVD}(b). 
The procedure is as follows.
A $g_\mu\times g_\nu$ matrix, $A\equiv\ee^*\cdot\xxi_{\mu\nu}$, can always be decomposed into $A=V\Sigma W^\dag$, with a $g_\mu\times g_\mu$ unitary matrix $V$, a $g_\nu\times g_\nu$ unitary matrix $W$, and a $g_\mu\times g_\nu$ matrix
\begin{align}
\Sigma&=\left(
\begin{array}{ccc|c}
\sqrt{\lambda_1} &        &                  &   \\
                 & \ddots &                  & 0 \\
                 &        & \sqrt{\lambda_r} & \\ \hline
                 & 0      &                  & 0
\end{array}
\right).
\end{align}
$\lambda_n$ is a positive number, and $r\le\min(g_\nu,g_\mu)$ is called the rank of $A$.
With this decomposition, the total transition matrix element, i.e., the trace of $\mathcal{M}_{\nu\mu}(\ee)$, is reduced to
\begin{align}
\sum_{bc}e^be^{c*}M_{\nu\mu}^{bc}&=\operatorname{Tr}A^\dag A=\sum_{n=1}^rM_{\nu\mu,n}(\ee),\label{eq:eeM_1}
\end{align}
where we have rewritten $\lambda_n$ as $M_{\nu\mu,n}(\ee)$.
One can see that the interband transition $\mu\to\nu$ is now decomposed into a smaller number of possible paths labeled by $n=1,2,\cdots,r$, each having a transition matrix element $M_{\nu\mu,n}(\ee)$.
It is straightforward to find its correction
\begin{align}
\Delta M_{\nu\mu,n}(\ee)&\equiv2\sqrt{\lambda_n}\operatorname{Re}[\ee\cdot(W^\dag C_{\nu\mu}V)_{nn}\Ee_0],
\end{align}
from
\begin{align}
\sum_{bcd}e^be^{c*}T_{\nu\mu}^{bcd}E_0^d
&=\sum_{n=1}^r\Delta M_{\nu\mu,n}(\ee).
\end{align}
The signular value decomposition is particularly useful for the shift vector.
One obtains
\begin{align}
\sum_{bc}e^be^{c*}S_{\nu\mu}^{bca}&=-\frac{\i}{2}\Tr\left[A^\dag(D^aA)-(D^aA^\dag)A\right]\nonumber\\
&=\sum_{n=1}^rM_{\nu\mu,n}(\ee)R_{\nu\mu,n}^a(\ee).\label{eq:eeS_1}
\end{align}
Here, $\Rr_{\nu\mu,n}(\ee)\equiv(\bar{\xxi}_{\nu\nu})_{nn}-(\bar{\xxi}_{\mu\mu})_{nn}$ is identified as the shift vector accompanying the $n$-th path, while
$\bar{\xxi}_{\nu\nu}\equiv W^\dag(\xxi_{\nu\nu}+\i\nabla_{\kk})W$ and
$\bar{\xxi}_{\mu\mu}\equiv V^\dag(\xxi_{\mu\mu}+\i\nabla_{\kk})V$ are Berry connections for the initial and final states, respectively.
Equation \eqref{eq:eeS_1} applied to Eq. \eqref{eq:Jinjene_0} lets us recognize $\hbar\Delta\omega_{\nu\mu,n}(\ee)\equiv-e\Ee_0\cdot\Rr_{\nu\mu,n}(\ee)$ as the field-induced energy shift for the $n$-th path.
In practice, it is more convenient to first calculate $\Dd_{\nu\mu}\equiv-e(\Ee_0\cdot\Dd_{\kk})\xxi_{\nu\mu}$ and $B\equiv\ee^*\cdot\Dd_{\mu\nu}$ for an arbitrary $\Ee_0$.
According to an identity
\begin{align}
\operatorname{Im}\Tr A^\dag B
&=\sum_{n=1}^rM_{\nu\mu}(\ee)\hbar\Delta\omega_{\nu\mu,n}(\ee)\nonumber\\
&=\sum_{n=1}^r\sqrt{\lambda_n}\operatorname{Im}(V^\dag BW)_{nn},\label{eq:eq_1}
\end{align}
we can express
\begin{align}
\hbar\Delta\omega_{\nu\mu,n}(\ee)&=\frac{1}{\sqrt{\lambda_n}}\operatorname{Im}(V^\dag BW)_{nn},
\end{align}
from which the shift vector follows:
\begin{align}
\Rr_{\nu\mu,n}(\ee)&=-\frac{\hbar}{e}\frac{\partial\Delta\omega_{\nu\mu,n}}{\partial\Ee_0}.
\end{align}

\end{document}